\documentclass[mathleft
]{an}
\usepackage{graphicx}
\usepackage{times}
\def\otres{[\ion{O}{iii}] $\lambda$ 5007 }

\def\aj{AJ}

\def\apj{ApJ}

\def\apjs{ApJS}

\def\aap{A\&A}
\def\aaps{A\&AS}
\def\mnras{MNRAS}

\def\nat{Nature}

\def\pasp{PASP}

\begin{document}

\Pagespan{789}{}
\Yearpublication{2005}%
\Yearsubmission{2006}%
\Month{11}%
\Volume{999}%
\Issue{88}%

\title{Relationship between the \otres line and 5GHz radio emission}

\author{A.Labiano\inst{1,2,3}
}
\titlerunning{\otres and radio emission}
\authorrunning{A. Labiano}
\institute{
Instituto de Estructura de la Materia, Serrano 121, 28006, Madrid, Spain
\and 
Kapteyn Astronomical Institute, Groningen, 9700 AV, The Netherlands
\and
Rochester Institute of Technology, Rochester, NY, 14623, USA}

\received{}
\accepted{}
\publonline{later}


\abstract{%
I have compiled observations of \otres line and 5 GHz radio emission for a large sample of GPS, CSS and FR sources. Several properties were studied and compared. The most relevant findings are that the FWHM and the luminosity of the \otres line are correlated with the size of the radio source. I present the data and discuss the correlations, with special focus on jet-host interaction, triggering and enhancing of \otres emission.}

\maketitle

\section{Introduction}

Current models for the evolution of powerful radio galaxies suggest that these sources propagate from the $\sim 10$ pc to Mpc scales at roughly constant velocity through an ambient medium which declines in density as $\rho(R) \propto R^{-2}$ while the sources decline in radio luminosity as $L_{rad} \propto R^{-0.5}$ (e.g., O'Dea 2002, and references therein). In this scenario, GPS would evolve into CSS and these into large -supergalactic sized- sources\footnote{Recent work added the High Frequency Peakers to the sequence, as possible progenitors of GPS (e.g., Orienti 2007 et al. and references therein).}. Such a scenario is consistent with the observed number densities of powerful radio sources as a function of linear size (e.g. O'Dea \& Baum 1997, Fanti et al. 2001). However, to match observations, the radio jets of the young sources must slow as they cross the host ISM and dim faster than predicted. The most likely mechanism to produce these effects is interaction of the radio source with the host environment.

Interaction was found (e.g., Labiano et al. 2008;  Holt et al. 2006; Axon et al. 2000). However, the studies focused in small samples or even just a few sources. Until now, there was no study of interaction of a large representative sample of GPS and CSS sources. Furthermore, the sparseness of samples studied did not allow a general study of the consequences of interaction on the models or the host.

In order to study the interaction between the gas clouds and the radio source I collected a sample of almost one hundred sources, including GPS and CSS galaxies and quasars, as well as FR sources. I study the properties of the \otres line and 5 GHz radio emission looking for traces of interaction and the mechanisms responsible for it.

\section{The sample}
\label{sec:sample}

I have compiled an representative sample of GPS, CSS and large sources with observations of the \otres $\lambda$ 5007 line and 5 GHz  radio emission. The sample consists of literature data (O'Dea 1998; Gelderman \& Whittle 1994, de Vries et al. 2000),the 2-Jy radio sources (Morganti et al. 1997; di Serego-Alighieri et al. 1994; Morganti et al. 1993; Tadhunter et al. 1993), STIS data (O'Dea et al. 2002; Labiano et al. 2005) and new Kitt Peak observations of three GPS radio galaxies: \object{0554-026}, \object{0941-080}, \object{1345+125}. The total number of sources included in the sample is 95 (21 GPS, 22 CSS, 52 large sources).


\section{Results and discussion}
\label{sec:results}

The following properties of the \otres line and 5 GHz radio emission were compared: FWHM, luminosity, asymmetry, kurtosis (\otres line), power, size and turnover frequency (radio emission) in the sample.

\begin{figure}
\centering
\includegraphics[width=\columnwidth]{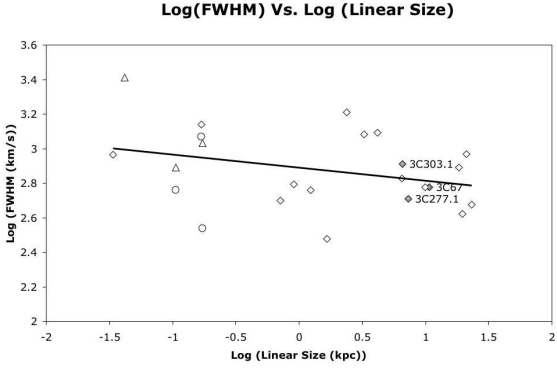}
\caption{Plot of the \otres FWHM of the sources in Gelderman \& Whittle 1994 (diamonds), de Vries et al. 2000 (triangles) our Kitt Peak (circles),  STIS (shaded diamonds) observations, and a linear fit to the data. The ionization and kinematics of \object{3C~67}, \object{3C~277.1} and \object{3C~303.1} are studied in Labiano et al. (2005) and O'Dea et al. (2002).  \label{FWHMLS}}
\end{figure}

The \otres FWHM shows no correlation with radio power, turnover frequency and \otres luminosity. Therefore, the shock velocity (closely related to the FWHM, see e.g., Bicknell et al. 1997) is independent of the strength radio source and does not affect the luminosity of the ionized gas or the radio spectral properties of the source. However, the data suggests a possible correlation between \otres FWHM and radio source size (Figure \ref{FWHMLS}) suggesting a possible deceleration of the jet as it crosses the host:

\begin{center}
log FWHM $\simeq  2.89(\pm0.04) - 0.08(\pm0.05) \times$ log LS  
\end{center}

The correlation is not very strong. However, the deceleration of the jet is required to predict observations (e.g., O'Dea 1998) and the latest models are predicting decelerations in the jet (see e.g., Kawakatu et al. 2009, in this volume). Unfortunately, there is not much data available of \otres FWHM.

\begin{figure}
\centering
\includegraphics[width=\columnwidth]{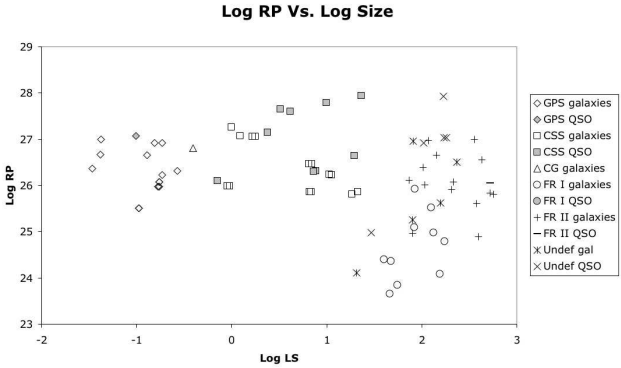}
\caption{Radio power at 5GHz versus linear size for GPS, CSS and large radio sources. \label{RPLS}}
\end{figure}

As expected (e.g., O'Dea \& Baum 1997, the sample also shows no correlation between radio power and radio size (Figure \ref{RPLS}), showing that GPS, CSS and FR2 sources are equally powerful (log Power$_{\mathrm{5 GHz}}\sim10^{26-27}$) while FR1 sources tend to be fainter. 

It is also clear that quasars tend to be brighter in \otres radio than galaxies, consistent with Unification scenarios: some of the \otres may be hidden by the torus in radio galaxies ()e.g. Hes et al. 1993). However, the difference in \otres emission could be due to selection effects: quasars are usually found at higher redshifts. The sample could be missing fainter quasars with luminosities similar to radio galaxies. 

\begin{figure}
\centering
\includegraphics[width=\columnwidth]{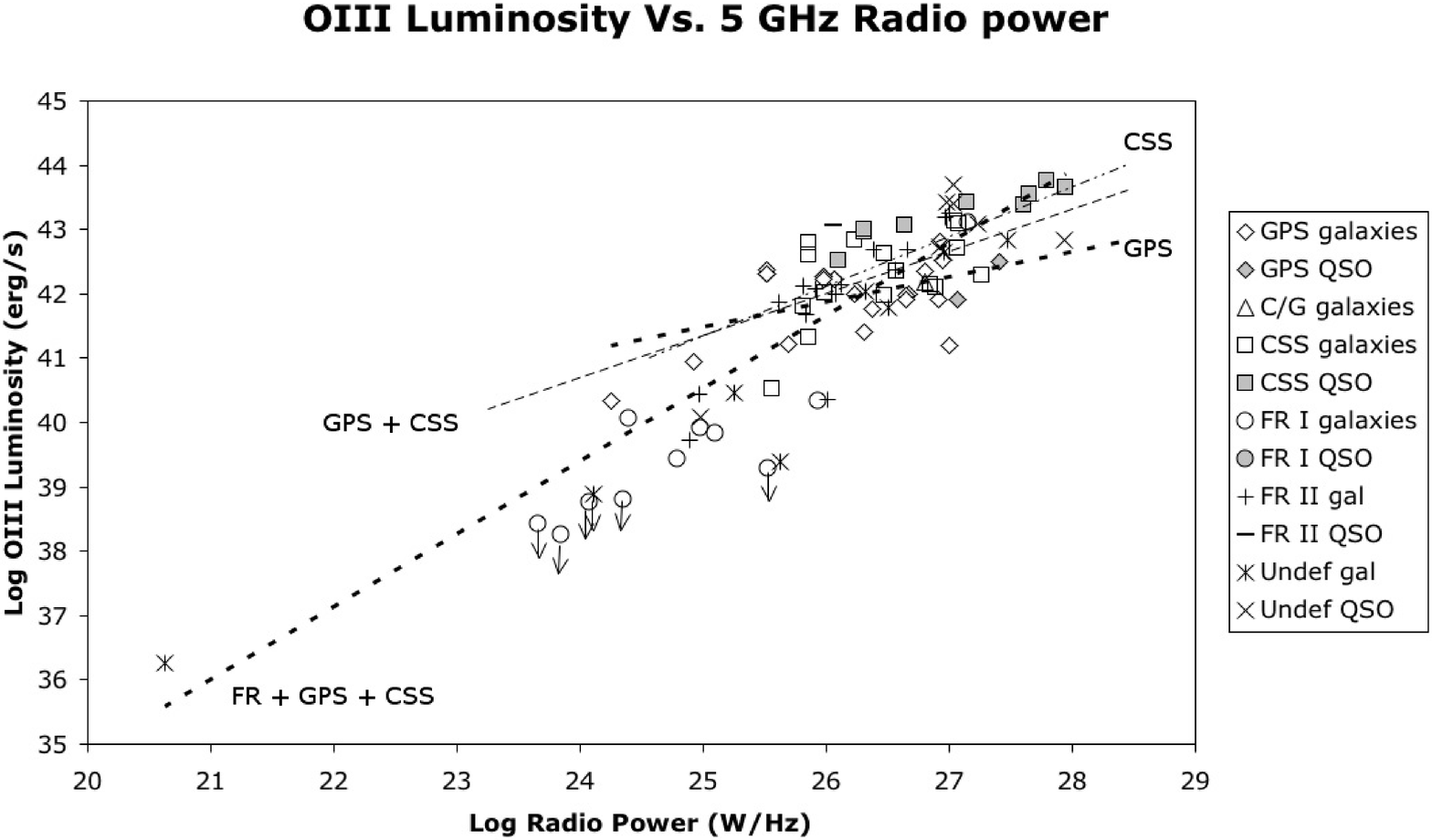}
\caption{Plot of the \otres luminosity versus radio power at 5GHz. Data for the CSS and GPS sources from Gelderman \& Whittle (1994); O'Dea (1998), de Vries et al. (2000), the 2-Jy sample and our Kitt Peak observations. Data for the large radio sources from the 2-Jy sample. The IDs of the sources have been updated with Zirbel \& Baum (1995). I use "C/G" to name those sources with no clear ID as CSS or GPS. \label{O3RP}}
\end{figure}

\begin{figure}
\centering
\includegraphics[width=\columnwidth]{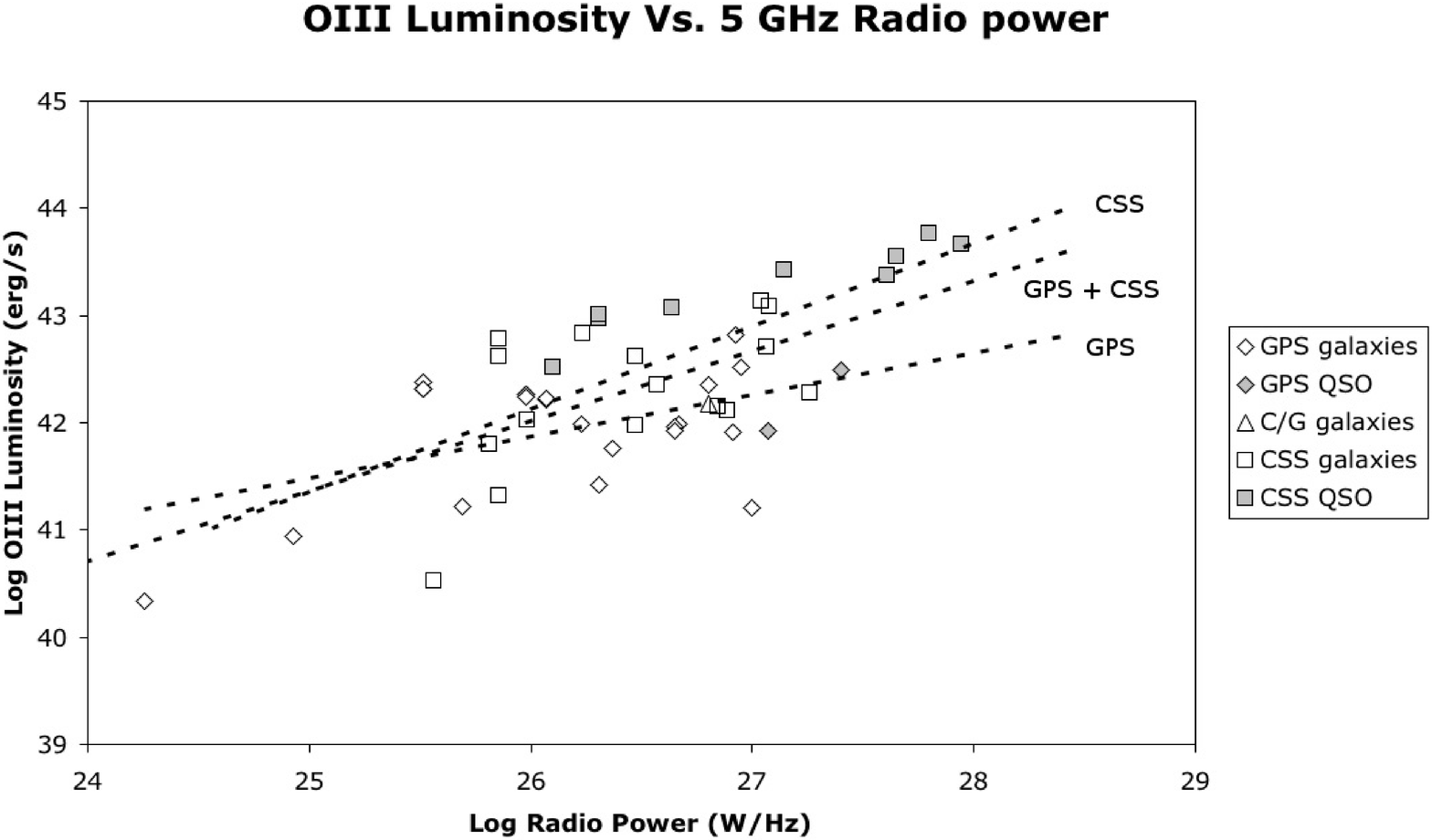}
\caption{As Figure \ref{O3RP}, showing only the GPS and CSS sources. \label{O3RPzoom}}
\end{figure}

The sample shows the known relation (e.g., Rawlings \& Saunders 1991;  Baum \& Heckman 1989) between \otres luminosity and radio power (Figures \ref{O3RP} and \ref{O3RPzoom}) for powerful (log Power$_{\mathrm{5 GHz}}> 10^{25}$) radio sources. This relation is usually explained as the AGN powering both the ionized gas and radio emission. For our sample, the different correlations are: 

\begin{flushleft} GPS: \end{flushleft}
\begin{center}  log L$_{[\ion{O}{iii}]}$ = 32($\pm$4) + 0.4($\pm$0.1) $\times$ log P$_{5GHz}$ \end{center} 
CSS:
\begin{center}  log L$_{[\ion{O}{iii}]}$ = 22($\pm$4) + 0.8($\pm$0.2) $\times$ log P$_{5GHz}$\end{center}  
GPS + CSS: 
\begin{center} log L$_{[\ion{O}{iii}]}$ = 25($\pm$3) + 0.7($\pm$0.1) $\times$ log P$_{5GHz}$ \end{center} 
GPS + CSS + large: 
\begin{center} log L$_{[\ion{O}{iii}]}$ = 12($\pm$2) + 1.13($\pm$0.07) $\times$ log P$_{5GHz}$ \end{center}

It should be noted that the correlation is not present in fainter radio sources (SDSS, Best 2008, in this volume), suggesting that a different mechanism may be producing the radio and \otres emission in these sources.

There is no evident relation between kurtosis or asymmetry of the \otres line with other properties of the line. Most of the sources have kurtosis lower than 1, suggesting the presence of broad wings in the gas. Whittle (1985) also finds the same for his sample of Seyfert galaxies. Most sources shows asymmetry values close to 0 in their \otres profile, suggesting the cocoon widens or narrows in a symmetric way  or there are no major variations between both sides. However, the ground spectra may lack of enough resolution to separate more complex structures that could change the profile of the line.

O'Dea (1998) discovered that in the Gelderman \& Whittle (1994) sample, GPS galaxies tend to have lower \otres luminosity than CSS galaxies. However, it was not clear if the trend would be followed by a larger sample with different selection criteria, what was the behaviour for quasars and large FR sources, or the possible consequences for general radio source evolution\footnote{A direct consequence is that it is more difficult to find optical counterparts of GPS than CSS.}. To asses these issues, I tested the trend with the current sample, which also includes GPS and CSS quasars and supergalactic-sized sources. I find that the GPS and CSS sources (galaxies and quasars) show a strong correlation between \otres line luminosity and size of the radio source\footnote{Note that the trend is also suggested by Figure \ref{O3RPzoom} and the different correlations between radio power and \otres luminosity for GPS and CSS sources.}:   

\begin{flushleft} GPS + CSS: \end{flushleft}
\begin{center} log L$_{[\ion{O}{iii}]}$ = 42.43($\pm$0.09) + 0.46($\pm$0.09) $\times$ log LS$_{5GHz}$ \end{center}
 
\begin{flushleft} GPS$^*$ + CSS:  \end{flushleft}
\begin{center} log L$_{[\ion{O}{iii}]}$ = 42.44($\pm$0.08) + 0.4($\pm$0.1) $\times$ log LS$_{5GHz}$ \end{center}

GPS$^*$ means the complete sample of GPS except the smallest source. 


In principle, this correlation could be due to the AGN enhancing both the radio and emission gas luminosities by photoionization. However, for the same radio power, small sources (GPS) are systematically fainter in \otres than larger (CSS) sources. 

\begin{figure}
\centering
\includegraphics[width=\columnwidth]{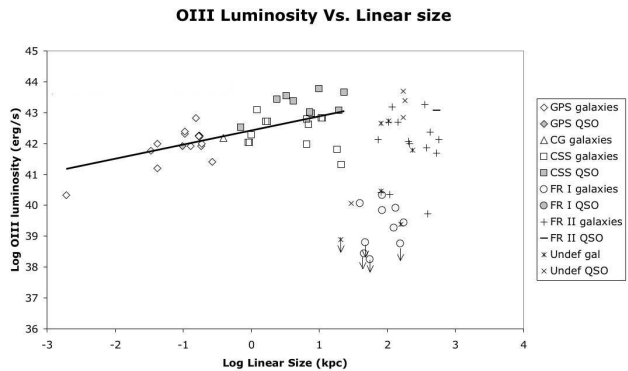}
\caption{Plot of the [\ion{O}{iii}] luminosity versus linear size of the radio source, showing the correlation for GPS and CSS sources. Data from the same references as figure \ref{O3RP}. \label{O3LS}}
\end{figure}

I propose a scenario where the expansion of the radio source through the host ISM is triggering and/or enhancing the \otres line emission through direct interaction. Some contribution from the AGN light must be present, however, AGN light alone, would not produce a correlation with size. Furthermore, the fact that the correlation disappears for supergalactic-sized sources supports this model: once the radio lobes leave the host, the \otres luminosity drops (Figure\ref{O3LS}). 

This scenario is also supported with previous observations finding evidence of strong interaction between the jet and surrounding ISM, as well as proof of shock-ionized \otres (e.g., Labiano et al. 2005; O'Dea et al. 2003). Jet-ISM interaction is also found through  \ion{H}{i} studies (e.g., Labiano et al. 2006; Holt et al. 2006, and references therein) and predicted by jet expansion models (e.g., Jeyakumar et al. 2005; Saxton et al. 2005)

Another interesting enhancing mechanism to consider is that the jet could enhance \otres, at least partly, through indirect mechanisms such as jet-induced star formation (Labiano et al. 2008, in prep.). However, new data are needed to study the possible contribution of recently formed stars. The hosts of GPS and CSS sources are usually elliptical galaxies so it is unlikely that the average/normal stellar population of the host has a strong to the \otres emission. These ``normal'' stars would however not create a correlation with radio jet size (and jet-induced stars would).

The correlation divided by source type is:\\

GPS:
\begin{center} log L$_{[\ion{O}{iii}]}$ = 42.8($\pm$0.2) + 0.8($\pm$0.2) $\times$ log LS$_{5GHz}$ \end{center}
 
GPS$^*$:
\begin{center} log L$_{[\ion{O}{iii}]}$ = 42.5($\pm$0.4) + 0.5($\pm$0.4) $\times$ log LS$_{5GHz}$ \end{center}

CSS:
\begin{center} log L$_{[\ion{O}{iii}]}$ = 42.6($\pm$0.2) + 0.2($\pm$0.3) $\times$ log LS$_{5GHz}$\end{center} 

The correlation tends to disappear when the sample is divided in different types of sources. This could be due to low statistics or to the smaller range of sizes covered by each type. GPS could show lower \otres due to high obscuration. However, it is more likely that young compact sources are too small to strongly affect their environment (this effect has also been observed in star formation histories of GPS hosts, Labiano et al. 2008). The UV luminosities of GPS seem to be as high as the CSS luminosities. Furthermore, the UV luminosity of GPS sources could be correlated with their radio power (Labiano et al. 2008. These two effects suggest that obscuration is not too strong in GPS sources or, at least, similar to CSS.


Concerning the overall scenario of radio source evolution, where GPS and CSS evolve into the large, supergalactic sized, sources, the visual inspection of Figure \ref{O3LS} suggests that CSS would evolve into FR2. Some authors also found a possible decreasing trend linking FR2 to FR1 (Best 2008, private communication) with increasing size. However, our sample needs more supergalactic-sized sources to address evolution beyond $\sim15-20$ kpc. Extensive discussions on the FR2 - FR1 connection can be found on the literature: see e.g., M{\"u}ller et al. 2004; Best et al. 2005; Wold et al. 2007 and references therein, or classical papers such as Baum et al. (1995) and Zirbel \& Baum (1995).

\section{Summary of main results and future work}

The aim of the project was to deepen our understanding of radio jet-host interaction in young AGN, studying the \otres line and 5 GHz radio emission properties of GPS and CSS sources. I have compiled a large ($\sim100$ sources), representative sample of GPS and CSS quasars, combined with FR 1 and FR 2, large sources (to help establish a time evolution line) from published data, as well as new data for three GPS sources.

The main result of the study is that the \otres emission is clearly enhanced by the jet expansion through the host ISM. This is consistent with previous observations as well as numerical models of jet expansion. However, further work is required (Labiano et al. 2008, in prep.):

\begin{itemize}
\item The supergalactic-size sample needs to be widen, to establish evolution when the radio lobes leave the host.
\item Most likely, the \otres emission is enhanced by the jet through shocks. However, it could also be enhanced by star formation induced by gas compressed by the shock, or a combination of both effects.
\item Study of star formation and AGN tracers (such as X-rays, 24 and 70 $\mu$m dust, line ratios and PAH, etc) to separate and evaluate the different mechanisms and contributions to the \otres emission.
\item Apply and improve jet expansion models to reproduce the results.

\end{itemize}

A parallel result to the \otres - radio size correlation is that it will be more difficult to find optical counterparts of GPS than CSS.

The data suggests a possible deceleration in the jet as it crosses the host ISM, which is required by most radio source evolution models. However, the correlation is too weak and a much wider sample, with higher resolution spectra, is needed.

The sample also reflects some already known results such as quasars showing lower \otres luminosity than radio galaxies, radio power correlated with  \otres luminosity, and radio power independent of the size of the radio source.

\acknowledgements
This research has made use of the NASA / IPAC Extragalactic Database (NED) which is operated by the Jet Propulsion Laboratory, California Institute of Technology, under contract with the National Aeronautics and Space Administration. 
I would like to thank C.P. O'Dea and P. D. Barthel for many fruitful comments and scientific discussions.

\end{document}